\documentclass{aastex}
\usepackage{emulateapj5}

\slugcomment{Accepted for the Astrophysical Journal Letters}

\shorttitle{Transits of GJ\,1214b}
\shortauthors{Sada et al.}

\begin{document}


\title{Recent Transits of the Super-Earth Exoplanet GJ\,1214b}


\author{Pedro~V.~Sada\altaffilmark{1,2}, Drake~Deming\altaffilmark{2,3}, Brian~Jackson\altaffilmark{2,3,4}, 
  Donald~E.~Jennings\altaffilmark{2,3}, Steven~W.~Peterson\altaffilmark{5}, Flynn~Haase\altaffilmark{5}, 
  Kevin Bays\altaffilmark{5}, Eamon O'Gorman\altaffilmark{6}, \&~Alan~Lundsford\altaffilmark{2,7}}


\altaffiltext{1}{Universidad de Monterrey, Monterrey, M\'exico}
\altaffiltext{2}{Visiting Astronomer, Kitt Peak National Observatory, National Optical Astronomy Observatory, 
  which is operated by the Association of Universities for Research in Astronomy under 
  cooperative agreement with the National Science Foundation}
\altaffiltext{3}{Planetary Systems Laboratory, NASA's Goddard Space Flight Center, 
  Greenbelt MD 20771}
\altaffiltext{4}{NASA Postdoctoral Fellow}
\altaffiltext{5}{Kitt Peak National Observatory, National Optical Astronomy Observatory, 
   which is operated by the Association of Universities for Research in Astronomy under 
  cooperative agreement with the National Science Foundation}
\altaffiltext{6} {Trinity College Dublin, Dublin 2, IRELAND}
\altaffiltext{7} {Catholic University of America, \& Planetary Systems Laboratory, 
   NASA's Goddard Space Flight Center, Greenbelt MD 20771}


\begin{abstract}
We report recent ground-based photometry of the transiting super-Earth
exoplanet GJ\,1214b at several wavelengths, including the infrared
near 1.25\,$\mu$m (J-band). We observed a J-band transit with the
FLAMINGOS infrared imager and the 2.1-meter telescope on Kitt Peak,
and we observed several optical transits using a 0.5-meter telescope
on Kitt Peak and the 0.36-meter Universidad de Monterrey Observatory
telescope. Our high-precision J-band observations exploit the
brightness of the M-dwarf host star at this infrared wavelength as
compared to the optical, as well as being significantly less affected
by stellar activity and limb darkening.  We fit the J-band transit to
obtain an independent determination of the planetary and stellar
radii.  Our radius for the planet ($2.61^{+0.30}_{-0.11} R_\oplus$) is
in excellent agreement with the discovery value reported by
Charbonneau et al. based on optical data.  We demonstrate that the
planetary radius is insensitive to degeneracies in the fitting
process.  We use all of our observations to improve the transit
ephemeris, finding $P=1.5804043\pm0.0000005$ days, and
$T_0=2454964.94390\pm0.00006$\,BJD.
\end{abstract}



\section{Introduction}

A current frontier in extrasolar planet research is the detection and
atmospheric characterization of super-Earth exoplanets orbiting nearby
M-dwarf stars \citep{charbdeming}.  The recent detection of GJ\,1214b
\citep{charb09} is a significant step toward this frontier.  GJ\,1214b
was found by the MEarth survey of nearby M-dwarfs
\citep{nutzman}, and it is the first super-Earth exoplanet discovered
by a ground-based photometric method.  The radius of GJ\,1214b derived
from the discovery observations ($2.68R_\oplus$, see Table~1) indicates
that the planet has a low molecular weight composition such as an
ocean planet, but alternative compositions remain possible
\citep{miller-ricci, rogers}. Inferences concerning the composition of
GJ\,1214b depend on a securely measured radius.  In this regard it is
valuable to measure the transit in the infrared (IR).  IR transit
observations are less sensitive to stellar activity and
limb-darkening, compared with optical wavelengths. Reduced sensitivity
to these effects helps establish a more robust planetary radius.
GJ\,1214 is also the target of several near-term observational
programs using both HST and Spitzer, so recent ground-based transit
observations can benefit the space-borne programs by improving the
precision of the transit ephemeris.

In this Letter we present results from high-precision photometry of a
transit of GJ\,1214b observed in the J-band (1.25\,$\mu$m), and
several less precise transits in the optical.  Our IR transit data
provide an independent check on the planetary radius derived by
\citet{charb09} from the optical discovery observations, and the
totality of our recent transit data allow us to significantly improve
the transit ephemeris.

\section{Observations and Photometry}

\subsection{Observations}

We observed transits of GJ1214b using the Universidad de Monterrey
Observatory (UDEM) telescope on UT dates April 29, June 6 \& 17,
2010. UDEM is a small private college observatory having Minor Planet
Center Code 720, and is located at 689 meters altitude in the suburbs
of Monterrey, M\'exico. The UDEM data were acquired using an
$I_c$-band filter on the 0.36-meter reflector, with a 1280x1024-pixel
CCD camera at 1.0 arc-sec per pixel scale.  We observed an additional
transit on UT May 29, 2010 using two telescopes on Kitt Peak.  We used
the 2.1-meter reflector with the FLAMINGOS 2048x2048-pixel infrared
imager \citep{elston}, and a J-band (1.25\,$\mu$m) filter, at 0.6
arc-sec per pixel scale. Following the conclusion of nightly public
programs, we have access to the 0.5-meter telescope at the Kitt Peak
Visitor Center (VCT).  Simultaneous with the 2.1-meter J-band
observations, we observed the May 29 transit using the VCT and a
3072x2048 CCD camera at 0.45 arcsec per pixel. The VCT observations
were planned to use the z-band filter, but due to a filter wheel error
these data were actually acquired in a B filter.  Observations at all
three telescopes used a defocus to improve the photometric precision,
and all used off-axis guiding to maintain pointing stability.
Exposure times were 30-seconds at the 2.1-meter, 60-seconds at the
VCT, and 120-seconds at UDEM.  All of the optical CCD exposures were
binned 2x2 to facilitate rapid readout.

Flat-field observations were acquired at all three observatories using
either twilight sky (UDEM and VCT) or a series of night-sky FLAMINGOS
exposures, incorporating pointing offsets to allow removal of stars
via a median filter.

\subsection{Photometry}

Subsequent to dark current subtraction and division by a flat-field
frame, we performed aperture photometry on the target star and
comparison stars.  The 20-arcmin field of FLAMINGOS provided 8
comparison stars of comparable IR brightness to GJ\,1214. We used
7-pixel circular aperture radii to measure the stars. We measured the
sky background surrounding each star using an annulus with an inner
radius of 9-pixels and an outer radius of 44-pixels. Normalizing
GJ\,1214 to the the comparison stars yielded a transit light curve
with an observed scatter that varied from 0.002 to 0.0015 as a
function of time, due to the decreasing airmass during the
observations.  We found that the best results were obtained by
averaging the ratios of GJ\,1214 to each comparison star.  This
produced smaller scatter than the method of ratioing GJ\,1214 to the
sum of all the comparison stars. We estimated an error for each
GJ\,1214 photometric point as the standard deviation of the ratio to
the individual comparison stars, divided by the square root of their
number (error of the mean).  The scatter in the photometry is
approximately 65\% greater than these estimates, so we increased our
error estimates by this factor to facilitate more accurate $\chi^2$
analyses (see below).

After normalizing to the comparison stars, the GJ\,1214 data clearly
reveal the transit but with low-order curvature in the out of transit
baseline. It is probable that this baseline curvature is caused by
differences in the effective wavelengths of the J-bandpass as a
function of stellar color, in combination with the
wavelength-variation in telluric water vapor absorption.  We account
for this baseline using a 4th-order polynominal fit, simultaneous with
the transit fit (see Sec.~3).  Photometry of the UDEM and VCT data
used similar procedures as for the FLAMINGOS data, except that the
baseline was modeled as a linear function of airmass.  Figure~1 shows
the 5 transits that we have observed.  Figure~2 replots the J-band
transit, showing both the raw photometry and baseline-removed photometry, and overplots a
theoretical transit curve based on the \citep{charb09} system parameters.

\section{Derivation of Stellar and Planetary Radii}

Our J-band transit data have greater photometric precision than our
optical data (see Figure~1).  Moreover, stellar limb darkening and
activity signatures are reduced at IR wavelengths compared to the
optical.  Therefore we use only our J-band data to fit for the
exoplanet and stellar radius, but we use all of our transit data to
update the transit ephemeris.

Given an estimate of the stellar mass, three unknowns (stellar and
planetary radii, and impact parameter) follow from three constraints
(transit depth and duration, and ingress/egress time). The solution is
only weakly affected by errors in the adopted stellar mass
\citep{brown}. Given also that the IR limb darkening is weak, we fix
both the stellar mass and limb darkening in our fitting for radii.  We
adopt the stellar mass (0.157 solar masses) given by \citet{charb09},
and the model-atmosphere-based square-root limb darkening coefficients for M-dwarf
stars derived by \citet{claret}.

We calculate transit curves numerically, using a tile-the-star
procedure with adaptive-mesh tiles to maximize the precision. The
transit curve that results from the \citet{charb09} system parameters,
and the \citet{claret} limb-darkening, is illustrated in the lower
panel of Figure~2.  For Figure~2, we overlaid this transit curve on
our data by first removing the baseline from the data. We then varied
the central transit time to find the best fit, not changing the depth
or shape of the curve.  It provides an excellent fit, note
particularly that it accounts for the sharp ingress/egress portions of
the data (Figure~2).  This agreement immediately suggests that the
\citet{charb09} radii are not signficantly affected by limb-darkening
errors, or stellar activity.  Note also that the transit curve still
contains significant curvature, unlike transit curves at far-IR
wavelengths \citep{richardson}, where stellar limb darkening is
virtually zero.  The linear limb darkening coefficient for this
M-dwarf star is about a factor of two lower in J-band than in I-band
\citep{claret}.

Given that the \citet{charb09} system parameters are very close to
best-fit values for our J-band data, we identify a refined solution as
follows.  We generate a 2-D grid of transit curves centered on the
\citet{charb09} values, as a function of the stellar radius and impact
parameter.  At each grid point, we find the best corresponding
exoplanet radius by scaling the depth of the transit curve and fitting
it to the raw photometry simultaneously with the baseline polynominal,
using linear regression. (Scaling the depth of the transit curve
relies on the excellent approximation that the transit depth is
primarily dependent on the ratio of planetary-to-stellar radius, and
not on the impact parameter.)  We also shift each transit curve in
time for the best fit.  At each grid point we calculate the $\chi^2$
of the best fit, and the location of the minimum $\chi^2$ gives our
global best-fit stellar radius, impact parameter, and planetary
radius.  We determine errors on these quantities via the range of
values contained within a $\chi^2$ contour of a given
significance. The results are given in Table~1, except for the
best-fit central transit time - that is included with the optical
transit times in Table 2.

To fit our four optical transits, we use the \citet{claret} limb
darkening coefficients to generate synthetic transit curves at the
optical wavelengths. We shift these in time, without varying their
depth, to find the best central time for each transit.  Those times
are listed in Table 2.  We combine our Table~2 transit times with the
individual transits given by \citet{charb09} to derive an updated
ephemeris, listed in Table~1.

\section{Results and Discussion}

Figure~3 shows our contours of $\chi^2$ for the fit to our J-band
transit, converted to standard deviations, and shown in the plane defined
by stellar radius and impact parameter.  Our $\chi^2$ minimum (square
on Figure~3) occurs at a smaller stellar radius and smaller impact
parameter than found by \citet{charb09}.  We do not regard this
difference as physically significant because: 1) it can reflect a
degeneracy in the fitting process, 2) the difference is not much
greater than the \citet{charb09} error range in combination with our
errors - the \citet{charb09} values are within our $2\sigma$ contour,
and 3) our radius for the planet is in excellent agreement with
\citet{charb09}.

The contours in Figure~3 trace a parabolic locus wherein a larger
stellar radius can trade-off versus a larger impact parameter (i.e.,
transit at higher latitude) to produce the same transit duration. In
the limit of a small planet and zero limb darkening, a chord at high
latitude across a large star would produce the same transit curve as a
chord at low latitude across a small star.  This mild degeneracy is
broken by the duration of ingress and egress, so that stellar radii
exceeding approximately $0.23R_{\odot}$ lie outside contours of
acceptable $\chi^2$.  We derive a value for the planetary radius
($R_p$) by examining the $R_p$ values at points lying within the
$1\sigma$ contour (that first contour is not illustrated on Figure~3).
This yields $R_p = 2.61^{+0.30}_{-0.11} R_\oplus$, where the
asymmetric error range reflects the fact that the radii are not
distributed symmetrically within the contour.  We also derived
best-fit values for the stellar radius and impact parameter, and these
are listed in Table~1 for comparison to the \citet{charb09} results.
Our error range for the impact parameter (a factor of two) is
relatively large, reflecting the degeneracy discussed above.
However, our radius for the planet has a much smaller fractional
variation within the $1\sigma$ contour, and is in excellent agreement
with \citet{charb09}. We conclude: 1) our value for the planetary
radius is relatively insensitive to the mild degeneracy between
stellar radius and impact parameter, and 2) the discovery value for
the radius of this super-Earth \citep{charb09} is unlikely to be
significantly affected by uncertainty in the optical limb darkening,
or affected by stellar activity.

\acknowledgements 

We are extremely grateful to Dr. Dick Joyce for considerable and
repeated help with FLAMINGOS, that contributed significantly to the
success of our J-band photometry at the 2-meter telescope. We are also
grateful to an anonymous referee for a swift and insightful review.

\clearpage



\begin{figure}
\epsscale{.80}
\plotone{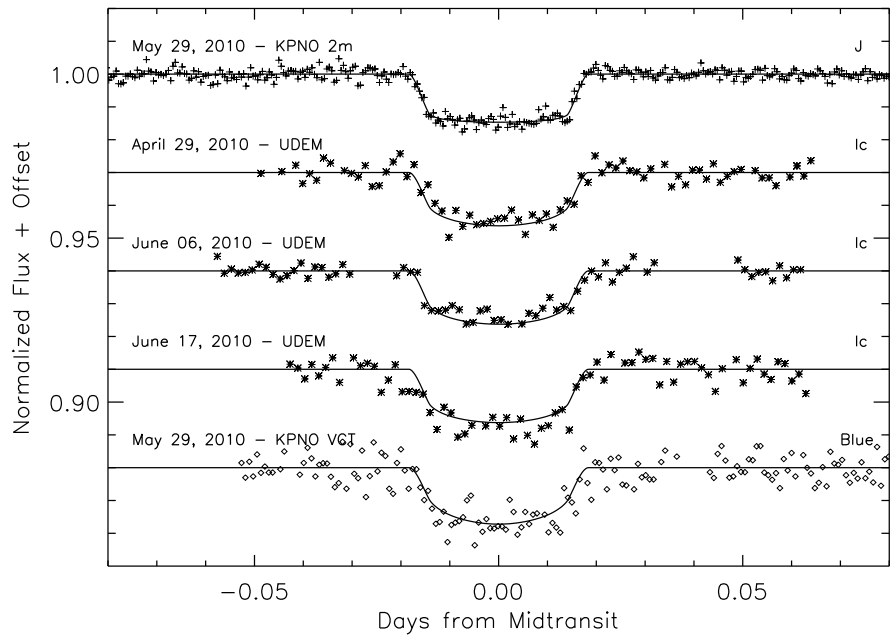}
\vspace{0.8in}
\caption{Photometry of GJ\,1214 during 5 transits using three different
telescope/wavelength combinations. Nominal transit curves are
overlaid, based on the \citet{charb09} system parameters, but shifted
to fit the best central transit time.
\label{fig1}}
\end{figure}

\clearpage

\begin{figure}
\epsscale{.50}
\plotone{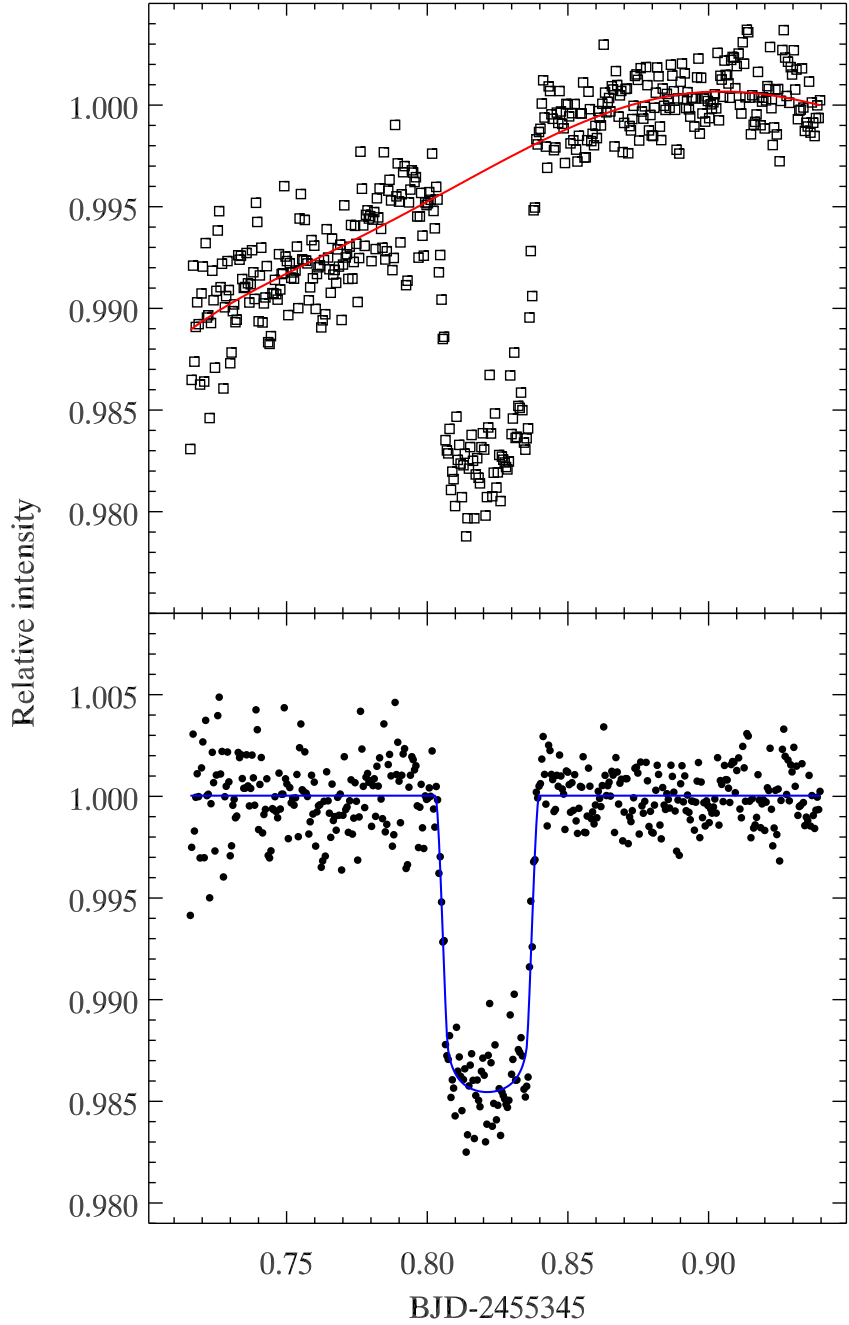}
\vspace{0.8in}
\caption{{\it Top Panel:} Raw J-band photometry of the GJ\,1214
transit observed from Kitt Peak using FLAMINGOS on May~29 UT. The red
line is a baseline fit using a 4th order polynominal, necessary due to
the different effective wavelengths of the target and comparison stars
when averaged over the filter bandpass.  {\it Bottom Panel:}
Photometry divided by the baseline, with a theoretical transit curve
(blue line) overlaid. The transit curve is based on IR limb darkening
for M-dwarf stars ($T=3000K$ \& $\log g = 5.0$) from \citet{claret},
and using the \citet{charb09} values for the system parameters.  It
was shifted in time to give the best fit, but was not altered in depth
or shape.  These J-band data are publically available in the NStED data base. 
\label{fig2}}
\end{figure}

\clearpage

\begin{figure}
\epsscale{.70}
\plotone{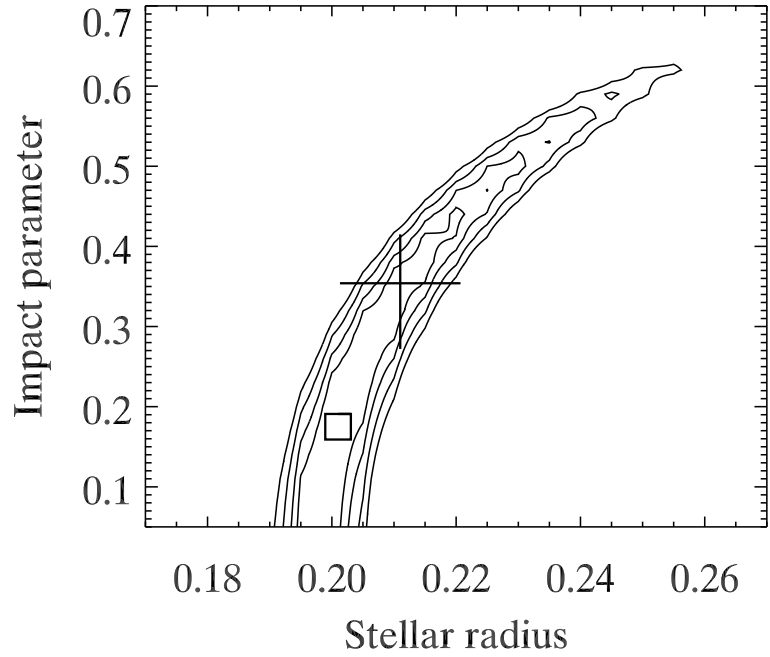}
\vspace{0.8in}
\caption{Contours of constant $\chi^2$ for the best fit to our J-band
transit, in the plane defined by the stellar radius and impact
parameter. The contours have been converted from $\chi^2$ to
statistical significance, and are plotted for 2, 3, 4, and 5$\sigma$
significance.  The cross shows the values and $\pm 1\sigma$ range from
\citet{charb09}, and the square indicates the position of our $\chi^2$
minimum.  Note the substantial range within the contours where the
same goodness of fit is obtained by trading a smaller stellar radius
for a lower crossing latitude (i.e., smaller impact parameter). This
is due to a mild degeneracy in the transit fit, as discussed in
Sec.~4.
\label{fig3}}
\end{figure}

\clearpage

\begin{table}
\begin{center}
\caption{Best-fit values for planetary and stellar radii, impact
parameter, and ephemeris parameters, compared to the results of \citet{charb09}.\label{tbl-1}}
\begin{tabular}{lll}
\tableline\tableline
Parameter & Best fitted value  & Charbonneau et al.(2009) values \\
\tableline
Planetary radius     & $2.61^{+0.30}_{-0.11}R_\oplus$      &   $2.678\pm0.13R_\oplus$  \\
Stellar radius       & $0.201^{+0.010}_{-0.005}R_\odot$    &   $0.2110\pm0.0097R_\odot$   \\
Impact parameter     & $0.175^{+0.181}_{-0.175}$           &   $0.354^{+0.061}_{-0.082}$  \\
Central transit time & $2454964.94390\pm0.00006$ (UTC-BJD) &   $2454983.9087558\pm0.0000901$  \\
Orbital period       & $1.5804043\pm0.0000005$\,days       &   $1.5803925\pm0.0000117$    \\
\tableline
\end{tabular}
\end{center}
Note: \citet{charb09} list only individual measured transit times, not
an averaged epoch.  We quote their most precise individual transit.
\end{table}

\begin{table}
\begin{center}
\caption{Times of GJ\,1214b transit center, given as UTC-based BJD. \label{tbl-2}}
\begin{tabular}{rccl}
\tableline\tableline
UT Date & Telescope & Filter  & BJD \& Error ($1\sigma$) \\
\tableline
April 29, 2010  & UDEM   & I$_c$   &   $2455315.79343\pm0.00042$  \\
May 29, 2010    & KP 2-m & J       &   $2455345.82126\pm0.00014$   \\
May 29, 2010    & KP VCT & B       &   $2455345.82133\pm0.00037$  \\
June 6, 2010    & UDEM   & I$_c$   &   $2455353.72332\pm0.00036$  \\
June 17, 210    & UDEM   & I$_c$   &   $2455364.78669\pm0.00029$  \\
\tableline
\end{tabular}
\end{center}
Note: Adding these transit times to those of \citet{charb09}, gives the best-fit ephemeris 
parameters (period and transit time) listed in Table~1.  
\end{table}



\clearpage





\begin{thebibliography}{}

\bibitem[Brown et al.(2001)] {brown} Brown,~T.~M., Charbonneau,~D., Gilliland,~R.~L., Noyes,~R.~W., \& 
    Burrows,~A., 2001, ApJ, 552, 699.

\bibitem[Charbonneau et al.(2009)] {charb09} Charbonneau,~D., \& 18 co-authors, 2009, Nature, 462, 891.

\bibitem[Charbonneau \& Deming(2007)] {charbdeming} Charbonneau,~D., \& Deming,~D., 2007, astro-ph/0706-1047.

\bibitem[Claret(1998)] {claret} Claret,~A., 1998, A\&A, 335,647.

\bibitem[Elston(1998)] {elston} Elston,~R., 1998, SPIE, 3354, 404.

\bibitem[Miller-Ricci \& Fortney(2010)] {miller-ricci} Miller-Ricci,~E. \& Fortney,~J.~J., 2010,
   ApJ, 716, L74.

\bibitem[Nutzman \& Charbonneau(2008)] {nutzman} Nutzman,~P. \& Charbonneau,~D. 2008, PASP, 120, 317.

\bibitem[Rogers \& Seager(2010)] {rogers} Rogers,~L.~A., \& Seager,~S., 2010, ApJ, 716, 1208.

\bibitem[Richardson et al.(2006)] {richardson} Richardson,~J.~L., Harrington,~J., Seager,~S. \&
    Deming,~D., 2006, ApJ, 649, 1043.

\end{thebibliography}
\end{document}